# Electron-Phonon Scattering in Planar MOSFETs: NEGF and Monte Carlo Methods

*Himadri S. Pal[1], Dmitri E. Nikonov[1,2], Raseong Kim[1,2], and Mark S. Lundstrom[1]*

[1]Department of Electrical and Computer Engineering, Purdue University, West Lafayette, Indiana 47906, USA

[2]Components Research, Intel Corporation, 2200 Mission College Blvd., Santa Clara, California 95052, USA

hpal@purdue.edu

## Abstract

A formalism for incorporating electron-phonon scattering into the nonequilibrium Green's function (NEGF) framework that is applicable to planar MOSFETs is presented. Restructuring the NEGF equations in terms of approximate summation of transverse momentum modes leads to a rigorous and efficient method of solution. This helps to drastically reduce the computational complexity, allowing treatment of both quantum mechanics and dissipative electron-phonon scattering processes for device sizes from nanometers to microns. The formalism is systematically benchmarked against Monte Carlo solutions of the classical Boltzmann transport for model potential profiles. Results show a remarkably close agreement between the two methods for variety of channel lengths and bias conditions, both for elastic and inelastic scattering processes.

## I. Introduction

Drift-diffusion simulation has been the basis of device transport modeling in most industry standard simulators, because it allows a fairly accurate analysis with low computational burden. The Boltzmann equation is often solved with the Monte Carlo approach to obtain parameters used in the drift-diffusion equations, to gain a deeper insight into the device operation, and explore new devices [1-3]. The Monte Carlo method can treat different scattering mechanisms, however it is a semiclassical method valid only in the long channel limit, with quantum mechanical effects included in a phenomenological manner [4]. Quantum mechanical formalisms, like the non-equilibrium Green's function (NEGF) technique, can incorporate the quantum effects such as confinement and tunneling in a rigorous manner. So far, however, it was mostly used for ballistic transport [5-7], which is a good approximation only for very small devices. As scaling of Si CMOS transistors approaches the 10nm size and alternative nanoscale devices are explored, there is an urgent need for an approach that treats both quantum mechanics and carrier scattering rigorously and which is universally valid independent of the device size.

Applying the NEGF formalism to solve for the ballistic transport in 2-D for a realistic device can be computationally intensive, and including scattering makes it almost intractable. Scattering can be incorporated in a phenomenological way by using Büttiker probes, but the connection between the self-energy and physical scattering mechanisms in this case is not straightforward and needs additional study [6, 8, 9]. Phonon scattering has been included within the self-consistent Born approximation for 1-D transport simulations in Si nanowires and carbon nanotubes [10-12]. The same technique has been used to simulate planar double-gate MOSFETs with extremely thin channels [13]. The large width of planar FETs requires solving for a large number of transverse modes and poses a computational challenge as compared to 1D structures, especially since scattering can couple different transverse modes. In this work, we demonstrate that the computational load is significantly reduced by using modified in/out-scattering and correlation functions that are integrated over the transverse dimension. We present results of rigorous benchmarking of the model with theory and Monte Carlo simulations, validating its use with Si channel devices. This allows us to rigorously treat the quantum effects as well as phonon scattering within a unified and computationally tractable framework for planar FETs.

The theoretical framework for including electron-phonon scattering in NEGF within the self-consistent first Born approximation is presented concisely in the next section, see [14] for a detailed derivation. The form of the in/out-scattering functions are derived for elastic (acoustic phonon via deformation potential) and isotropic inelastic (non-polar optical phonon) scattering. By elastic scattering we understand processes with little change in the carrier energy; the phase is in general randomized in such processes. In the derivation we make and justify an approximation of independence of Green's function on the transverse momentum. These have been validated by establishing their connections with scattering rates used in classical Boltzmann transport theory [14]. The appropriate conditions that allow the integration over the unconfined transverse dimension are then established. Section III describes the implementation of the approach in the NEGF based device simulator, nanoMOS [15], and the benchmarking results are presented in Sec. IV. The transport is solved for a fixed (non self-consistent) conduction band profile with both NEGF and Monte Carlo simulations, using one band and one phonon mode. A linear and a parabolic conduction band profile are used for the benchmarking, to mimic a resistive element and a transistor channel. Comparison results for both elastic and inelastic scattering process are presented, followed by the self-consistent simulations in Sec. V and conclusions in Sec. VI.

## II. Formalism

*II.1. General formalism*

According to the Keldysh non-equilibrium Green's function (NEGF) technique, the device is described by a Hamiltonian $H$, and the self-energy $\Sigma = \Sigma_c + \Sigma_s$ represent the effects due to semi-infinite source/drain contacts $(\Sigma_c)$ and all mechanisms of scattering $(\Sigma_s)$ [6, 7, 16]. If these are known, the retarded Green's function for the device is obtained as

$$G(E) = \left[\left(E + i\eta^+\right)I - H - \Sigma(E)\right]^{-1}, \qquad (1)$$

where $\eta^+$ is an infinitesimal positive value and $I$ the identity matrix. Any potential variation within the device is implicitly included in the Hamiltonian itself. The in/out scattering functions $\Sigma^{in/out}$ (they have contributions from contacts as well as scattering too) can be used to obtain the electron and hole correlation functions as $G^{n/p}(E) = G\Sigma^{in/out}G^\dagger$. The charge distribution within the device is proportional to the main diagonal of these correlation functions while the upper and lower diagonals correspond to the current distribution. The scattering component of $\Sigma^{in/out}$ are often dependent on $G^{n/p}$, and hence a self-consistent calculation is required to solve for the charge and current in the device.

Phonon scattering [7,14,16] is described by $M_q$ - the coupling strength of electrons with phonons in mode $q$, $b_q^\dagger, b_q$ - the creation and annihilation operators, and the half-amplitude for one phonon in the total device volume $V$ with density $\rho$

$$a_q = \sqrt{\frac{\hbar}{2\rho V \omega_q}}. \qquad (2)$$

The averages of the operator products in a reservoir at thermal equilibrium are given by $\langle b_q^\dagger b_{q'} \rangle = \delta_{qq'} n_q$ and $\langle b_q b_{q'}^\dagger \rangle = \delta_{qq'}(n_q + 1)$ (averages of all other products are zero). Here $n_q$ is the phonon occupation number of mode $q$, and is governed by Bose-Einstein statistics assuming an equilibrium phonon population.

$$\Sigma^{in}(r_1, r_2, E) = D(r_1, r_2, E)(n_q + 1)G^n(r_1, r_2, E + \hbar\omega_q) + D^*(r_1, r_2, E)n_q G^n(r_1, r_2, E - \hbar\omega_q), \qquad (3)$$

$$\Sigma^{out}(r_1, r_2, E) = D^*(r_1, r_2, E)(n_q + 1)G^p(r_1, r_2, E - \hbar\omega_q) + D(r_1, r_2, E)n_q G^p(r_1, r_2, E + \hbar\omega_q). \qquad (4)$$

The first and second terms in (3) and (4) denote phonon emission and absorption respectively. The coupling strength is incorporated in the electron-phonon coupling operator $(D)$, which operates on the terms on its right, and is given by

$$D(r_1, r_2, E) = \sum_q |M_q|^2 a_q^2 \exp(iq(r_1 - r_2)). \qquad (5)$$

*II.2. Basis transformation*

The above formalism is applied to the case of a planar MOSFET. The directional convention is adopted for the nanoMOS simulator [15]: the transport direction is designated as x, the confined direction perpendicular to the gate is z, and the unconfined direction along the width of the devices (considered infinite) is y. The grid sizes in these directions are $a_x, a_y, a_z$. Functions of spatial coordinates need to be transformed to the new basis using the eigenfunctions – plane waves along y-axis

$$\psi_{kt}(y) = N_y^{-1/2} \exp(ik_t y), \tag{6}$$

where $k_t$ is y-directed transverse momentum, and $N_x, N_y, N_z$ are the numbers of grid cells along the length $L_x, L_y, L_z$ in each direction used for normalization. The eigenfunctions along the confined direction $\psi_\alpha(z)$ are obtained by solving the Schrodinger equation with the confining potential, where index $\alpha$ corresponds to subbands. We assume that different subbands and states with different momenta are uncorrelated, i.e. their relative phases are random. In this case the off-diagonal elements between these states in the Green's functions and self energies vanish. This is not in contradiction with coupling of the modes by transfer of population in scattering processes. After the transformation to the new basis, the

$$\Sigma^{in}(x_1, x_2, \beta, k_t, E) = \sum_{ql, qt, \alpha} |M_q|^2 a_q^2 (n_q + 1) \exp(iq_l(x_1 - x_2)) \\ \times G^n(x_1, x_2, \alpha, k_t + q_t, E + \hbar\omega_q) F(\beta, \alpha) + abs. \tag{7}$$

where $q_l$ and $q_t$ are the longitudinal (x) and transverse (y) projection of the phonon wavevector respectively, and '*abs.*' denotes the absorption terms. So far the self-energy (7) is general and does not contain assumptions of isotropic scattering. A similar expression with substitution of the hole correlation function and the change of sign of the energy yields the out-scattering function $\Sigma^{out}$. The form-factor [17] for scattering $(F)$ accounts for all possible transitions between subbands $\beta$ and $\alpha$.

$$F(\beta,\alpha) = \sum_z \psi_\beta^*(z)\psi_\alpha^*(z)\psi_\beta(z)\psi_\alpha(z). \tag{8}$$

*II.3. Optical phonons*

The in/out scattering functions can be simplified for isotropic scattering with phonons of constant energy (with constant matrix element $|M_q| \approx D_0$, the phonon frequency $\omega_q \approx \omega_0$, and the phonon occupation number $n_q \approx n_0$), which is satisfied by optical phonons and deformation potential interaction in non-polar crystals. Since $\sum_q \exp(iqr) = N\delta(r)$ for summation over the Brillouin zone, the in/out-scattering functions are diagonal in this case [14]:

$$\Sigma^{in}(x_1,x_1,\beta,k_t,E) = K_o(n_0+1)\frac{a_y}{L_y}\sum_{qt,\alpha} G^n(x_1,x_1,\alpha,k_t+q_t,E+\hbar\omega_0)F(\beta,\alpha) + abs. \tag{9}$$

This results in a diagonal self-energy term, permitting the use of recursive inversion algorithms (applicable for tri-diagonal matrices) for computing the Greens function [17]. The optical phonon coupling constant $(K_o)$ is defined as

$$K_o = \frac{\hbar D_0^2}{2\rho\omega_0 a_x a_y a_z}. \tag{10}$$

*II.4. Acoustic phonons*

The in/out scattering functions can be drastically simplified in the case of acoustic phonon scattering by deformation potential in non-polar crystals. Acoustic phonon energy is relatively small and this can be considered as elastic scattering, and the phonon occupation number is typically very high $\left(n_q \approx k_B T/\hbar \omega_q \gg 1\right)$. For such low energy phonons, we can use the first linear term of expansion of energy and coupling matrix element over momentum ($\omega_q = v_a q$ and $|M_q| = D_a q$) [18]. Since the summation over longitudinal and transverse phonon momentum does not involve any electron momentum, the in-scattering function is given by [14]:

$$\Sigma^{in/out}(x_1, x_1, \beta, k_t, E) = K_a \frac{a_y}{L_y} \sum_\alpha G^{n/p}(x_1, x_1, \alpha, k_t, E) F(\beta, \alpha) \tag{11}$$

where the acoustic phonon coupling constant $(K_a)$ is [18]:

$$K_a = \frac{D_a^2 k_B T}{\rho v_a^2 a_x a_y a_z} \tag{12}$$

The in/out scattering functions are diagonal in this case, and there is no dependence on the transverse phonon momentum.

*II.5. Approximate summation over the unconfined transverse dimension*

Here we make an approximation under which the simulation complexity can be greatly reduced. We would like to eliminate the need to treat each value of transverse momentum separately. This approximation is often done for ballistic transport simulation [15] and is valid for parabolic bands and constant mass across the device, when the Hamiltonian is separable into independent longitudinal and transverse part, and there is no transverse variation of bias $\left(H_t(k_t) = \hbar^2 k_t^2/2m_t = E_t\right)$ or contact self-energy. Here we expand this approximation to cases with scattering. The Green's function is

$$G(E_l, E_t) = \left[ \left(E_l + i\eta^+\right)I - H_l - \Sigma_c(E_l) - \Sigma_s(E_l, E_t) \right]^{-1}. \tag{13}$$

The specific dependence on the unconfined transverse dimension arises from the scattering self-energy $(\Sigma_s)$, so if this dependence is eliminated, the transverse momentum summation can be resolved analytically. We define the electron and hole correlation functions summed over the unconfined transverse dimension as

$$\tilde{G}^{n/p}(E_l) = \frac{a_y}{L_y} \sum_{kt} G^{n/p}(k_t, E_l + E_t). \tag{14}$$

And similar expressions for the summed in- and out-scattering functions $\tilde{\Sigma}^{in/out}(E_l)$. Once we justify a posteriori the assumption that the scattering self-energy and thus the Green's function are independent of the transverse energy, we will prove that their summed versions are equal to themselves $\tilde{G}(E_l) = G(E_l)$ and $\tilde{\Sigma}(E_l) = \Sigma(E_l)$. Therefore the relation of the correlation functions and scattering functions is still valid.

$$\tilde{\Sigma}^{in/out}(E_l) = G(E_l)\tilde{G}^{n/p}(E_l)G^+(E_l). \tag{15}$$

For acoustic phonons, the in/out-scattering functions are proportional to the correlation functions (11), and have no explicit dependence on transverse momentum. The summation is trivial.

$$\tilde{\Sigma}^{in/out}(x_1, x_1, \beta, E_l) = K_a \sum_{\alpha} \tilde{G}^{n/p}(x_1, x_1, \alpha, E_l) F(\beta, \alpha) \tag{16}$$

For optical phonons, the summation is performed by introducing a change of variable $(k_t + q_t \to k_t)$ in the in/out-scattering functions of (9). Here we make the approximation that the longitudinal energy after scattering does not depend on the transverse energy. This approximation works best when the range of interest of longitudinal energies exceeds the phonon energy.

$$\tilde{\Sigma}^{in}(x_1, x_1, \beta, E_l) = K_o(n_0 + 1) \sum_\alpha \tilde{G}^n(x_1, x_1, \alpha, E_l + \hbar\omega_0) F(\beta, \alpha)$$
$$+ K_o n_0 \sum_\alpha \tilde{G}^n(x_1, x_1, \alpha, E_l - \hbar\omega_0) F(\beta, \alpha)$$
(17)

$$\tilde{\Sigma}^{out}(x_1, x_1, \beta, E_l) = K_o(n_0 + 1) \sum_\alpha \tilde{G}^p(x_1, x_1, \alpha, E_l - \hbar\omega_0) F(\beta, \alpha)$$
$$+ K_o n_0 \sum_\alpha \tilde{G}^p(x_1, x_1, \alpha, E_l + \hbar\omega_0) F(\beta, \alpha)$$
(18)

Thus, under the above assumptions we show that the scattering functions and self-energy are independent of the transverse energy. Therefore the Green's function is independent too. The corresponding electron/hole density and current can be computed as transverse integrated terms with only the longitudinal energy to be considered in the simulations.

*II.6. Self consistent loop*

When electron-phonon scattering is included, an added complexity is introduced as the in/out scattering energies and self energy are dependent on electron/hole density, making it a non-linear system. The calculation for the correlation functions $\left(\tilde{G}^{n/p}\right)$ has to be computed self-consistently with the in/out-scattering functions $\left(\tilde{\Sigma}^{in/out}\right)$. Equations (16)-(18) are used to compute $\tilde{\Sigma}^{in/out}$ from previous iteration values of $\tilde{G}^{n/p}$ (The initial guess is from a ballistic calculation). In general, $im(\tilde{\Sigma}_s) = -\frac{1}{2}(\tilde{\Sigma}^{in} + \tilde{\Sigma}^{out})$, and $re(\tilde{\Sigma}_s)$ is obtained from the Hilbert transform of $im(\tilde{\Sigma}_s)$. For elastic scattering with a constant coupling, it can be shown that $re(\tilde{\Sigma}_s)$ also has the same form as (16), so $\tilde{\Sigma}_s(m, E_l) = K_a \sum_\alpha G(\alpha, E_l) F(m, \alpha)$. The $Re(\tilde{\Sigma}_s)$ is included in our simulations for the sake of completeness, but it does not have an appreciable effect on the device characteristics. The Greens function is then computed using (1) and the electron/hole density $\left(\tilde{G}^{n/p}(E) = G\tilde{\Sigma}^{in/out}G^\dagger\right)$ are updated accordingly, and the loop is repeated till self-consistency is achieved.

*II.7. Relation to Büttiker probes*

We note that the method described here is essentially different from the Büttiker probe method [6]. That method introduces coupling to a fictitious reservoir at every grid point. The coupling to the reservoir is proportional to the expected scattering rate. The method has adjustable parameters - Fermi levels of each reservoir. They are calculated in a self-consistent loop to insure that the total flux of particle from the reservoir for all energies is equal to zero. The self energy in the Büttiker probe method does not depend on the occupation numbers of electrons and holes, and does not have a room to account for phonon energy. We believe that out method is more rigorous and physically justified.

## III. Implementation

Electron-phonon scattering has been implemented within the NEGF framework in the effective mass based device simulator nanoMOS [15, 19-21]. NanoMOS is deployed for public use at www.nanoHUB.org, and can simulate ultra-thin body double gate MOSFETs using drift-diffusion and ballistic NEGF transport. For quantum transport calculations, the 2D electrons in the channel layer are separated into confined subbands (decoupled mode-space approach), and transport in each mode is computed using 1-D NEGF separately [22]. The confined modes are uncoupled in the ballistic simulations, and coupled through phonons in the scattering calculations. The potential drop in the device is updated self-consistently using a 2D Poisson solver. Simulations involving scattering start with a ballistic calculation of the electron and hole correlation functions, $\tilde{G}^{n/p}$, which serves as the initial guess for computing the in/out-scattering functions. The coupling constants are obtained from deformation potentials and other material parameters, no fitting parameters are involved in the process. The in/out scattering functions and self energies are diagonal for the electron-phonon interactions considered, allowing the use of fast recursive algorithms to compute the Green's function instead of full matrix inversion. A self-consistent loop is implemented for inelastic scattering which monitors the convergence of $G$, $\tilde{G}^{n/p}$ and $\Sigma$, $\Sigma^{in/out}$, and another one for elastic scattering. Compared to the ballistic simulations, the two self-consistent loops are the only significant addition in terms of computational complexity. A typical simulation including scattering is 20~30 times slower than a ballistic simulation due to these self-consistent loops.

## IV. Benchmarking

*IV.1. Elastic Scattering*

A very thin channel double-gate device is used as a model for benchmarking the phonon scattering formalism. The transport in the channel region is decoupled from the electrostatics for benchmarking. A small constant electric field is applied to the channel (fixed linear conduction band profile), while the source/drain is assumed to be in equilibrium. Si is used as the channel material, and only the lowest unprimed X valleys are assumed to be occupied. The elastic scattering parameter used is that for intra-subband acoustic phonon scattering in bulk silicon [2]. Figure 1a shows the energy resolved current distribution along this resistive channel and the conduction band profile for an example simulation. There is no energy relaxation involved as only elastic scattering is the only scattering mechanism here. The energy resolved transmission and current for a ballistic channel and with scattering are shown in fig. 1b. Transmission is unity above the conduction band in the ballistic case as expected. Scattering reduces the transmission from this ideal limit, and the current also reduces accordingly.

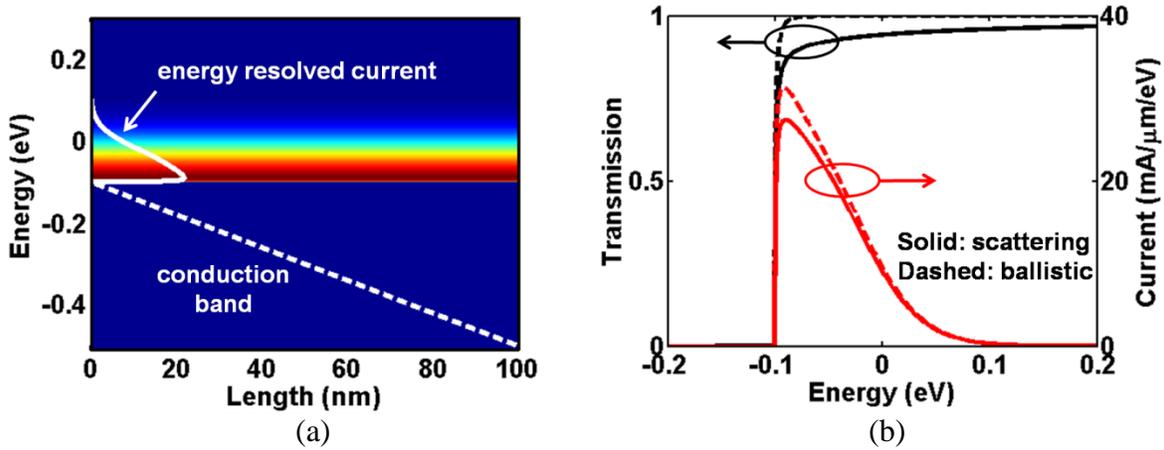

Fig 1: (a) Energy resolved current image with conduction band vs L, and (b) transmission and current distribution vs. energy, for a linear potential with ballistic transport, and acoustic phonon scattering.

The acoustic phonon scattering formalism is benchmarked with analytical calculations and Monte Carlo simulations. Although the Monte Carlo simulator is classical, it can accurately model transport for the linear conduction band assumed here, except for extremely scaled channels where quantum effects (tunneling, reflection of electron waves, metal induced gap states) cannot be ignored. The Monte Carlo simulator takes the incident-flux approach, assuming non-degenerate contacts [18]. In the Monte Carlo simulations as well as in the NEGF cases compared with them we set the Boltzmann distribution of carriers over energy. The simulation begins with a carrier selected at random from the flux injected from the source contact. The injected carrier is accelerated by the electric field along the transport direction during the free flight between scattering events, and the momentum is randomized in the two-dimensional momentum space after each scattering event. The carrier is followed until it comes back to the source or reaches the drain contact. We assume that the carriers that reach contacts simply exit the channel. When the carrier injection from the drain contact is suppressed, the transmission $T$ can be calculated by dividing the number of carriers reaching the drain contact by the total number of carriers injected from the source contact. The source Fermi level is aligned to the conduction band at the source, while the drain Fermi level is taken to be low enough to suppress any injection from the drain. A non-degenerate carrier statistics is assumed, as it makes the verification with an analytical calculation easier.

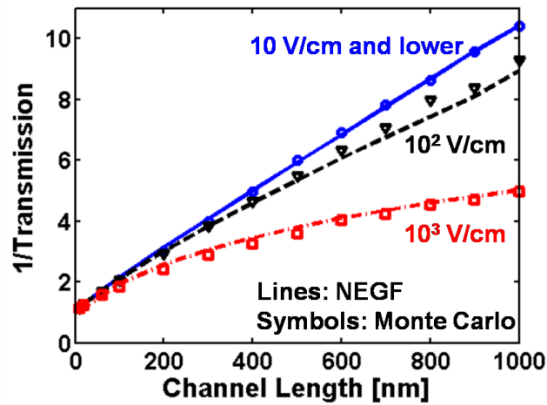

Fig. 2: Plot of transmission inverse vs. channel length for linear conduction band with varying electric fields obtained from Monte Carlo and NEGF simulations. The plot is linear for low electric field (< 10 V/cm), with a slope that's inverse of the mean free path $\lambda_0 = 104$nm for intra-subband acoustic phonons.

The transmission $T$ in a channel under low electric field can be expressed as $T = \lambda_0/(\lambda_0 + L)$, where $\lambda_0$ is the average backscattering mean free path and $L$ is the channel length [23]. This holds when the field is low enough to maintain near-equilibrium transport inside the channel. The inverse of transmission varies linearly with channel length $(1/T = 1 + L/\lambda_0)$, and its slope can be used to extract the mean free path. Figure 2 shows plots of $1/T$ vs. $L$ from Monte Carlo and NEGF simulations for varying channel electric fields. The transmission is averaged over the energy. The plots exhibit a linear behavior at low field, and deviates at higher field where the linear relationship doesn't hold. Under high electric field conditions, the transmission increases monotonically with increasing field [24]. The scattering rate $(1/\tau)$ and mobility $(\mu = q\tau/m)$ for acoustic phonon scattering in a 2-D electron gas is governed by material parameters like deformation potential and the 2-D density of states [23]. The mean free path $(\lambda_0)$ is linearly proportional to the mobility $(\mu = q\lambda_0 v_t/(2k_B T_L))$, which is used to compute $\lambda_0$ analytically. For a Si channel with 3nm thickness in the confined dimension, the analytical value of $\lambda_0$ is 104 nm for the acoustic deformation potential of 9.5 eV. The $\lambda_0$ extracted from Monte Carlo and NEGF simulations of the slope of $1/T$ vs. $L$ plots at low field closely match this analytically computed value.

According to Natori's theory of the ballistic MOSFET [25], the current is determined by the positive (source to drain) and negative going fluxes at the top of the barrier, and is expressed as $I_B = I^+ - I^- = qWn^+v^+ - qWn^-v^-$. Here, $n^+ = (N_{2D}/2)\mathsf{F}_0(\eta_F)$ is the carrier population of the positive $k$-states given by the effective 2D density of states $N_{2D} = g_v m_{DOS} k_B T_L / (\pi \hbar^2)$, where $g_v$ is the valley degeneracy factor, and $\eta_F = (E_F - E_C)/k_B T_L$ [26]. The injection velocity $v^+ = \sqrt{2k_B T_L / \pi m_y}\, \mathsf{F}_{1/2}(\eta_F)/\mathsf{F}_0(\eta_F)$ is the average velocity of the positive going flux. The Fermi-Dirac integrals ($\mathsf{F}$) of order 0 and 1/2 result from analytical integration over transverse modes. The corresponding negative-going flux and velocity are obtained by replacing $\eta_F$ with $\eta_F - qV_D/k_B T_L$. The current in the presence of scattering is simply reduced by the transmission $T = \lambda_0/(\lambda_0 + L)$, so that $I = T \times I_B$. For $L \ll \lambda_0$ transport is ballistic and $T = 1$, while for $L \gg \lambda_0$ transport is diffusive with $I = (\lambda_0/L)I_B$.

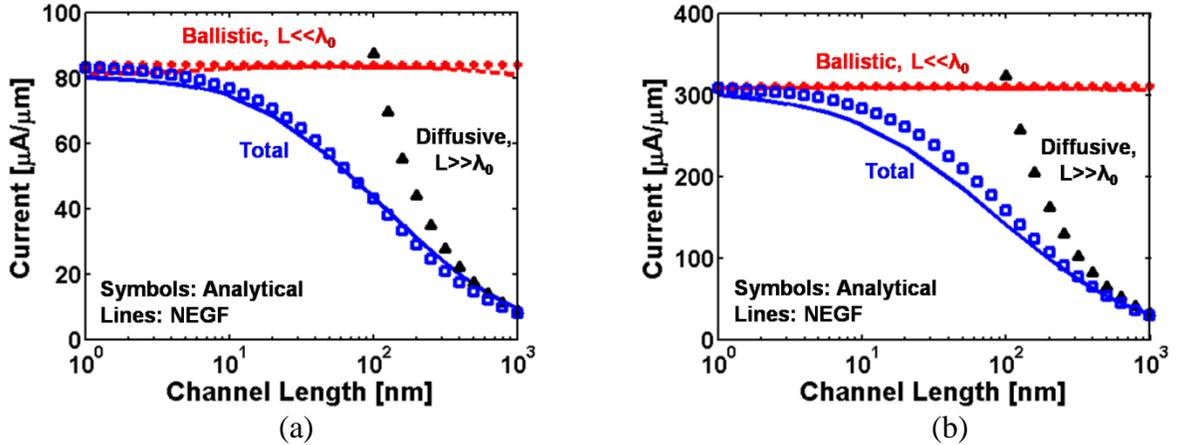

Fig. 3: Current per unit width vs. channel length for linear potential with zero electric field, obtained analytically and by NEGF simulations. (a) Low drain bias (10 mV) with injection from both source and drain and (b) high drain bias where injection from the drain end is suppressed.

Figure 3 plots the variation of current with channel length obtained using the above analytical calculation, and from NEGF simulations. A flat conduction band profile (zero electric field) is used in fig 3(a) with a low drain bias of 10 mV, so carriers are injected from both source and drain junctions. The results in fig. 3(b) are obtained similarly, but with a high drain bias such that any drain side injection is suppressed. The analytical results are plotted for the total current, as well as the currents with the ballistic and diffusive approximation. The ballistic NEGF simulations as well as those with electron-phonon scattering match very closely with the analytical calculations. The current computed using NEGF with scattering varies seamlessly from a ballistic limit (independent of $L$) at low $L$ to the diffusive limit ($1/L$ dependence) at large $L$. We observe that the ballistic approximation is valid for $L < 10 nm$, the diffusive limit is valid for $L > 300 nm$, while the full quantum result with scattering is universally valid for all channel lengths.

The scattering at the $k_B T_L$-layer near the top of the barrier determines the transmission and steady-state drain current of a MOSFET [27]. This is the length where the voltage drops by about one $k_B T_L$, and is of the order of one mean free path for well-designed devices. It is imperative to benchmark the phonon scattering model at the electric field magnitudes relevant to the $k_B T_L$-layer of a nanoscale MOSFET. Figure 4 plots the transmission as a function of the channel length for low and high electric fields. The transmission monotonically increases with electric field, and goes to one for fields exceeding $10^6$ V/cm. Transmission decreases with increasing channel length as expected, and shows a close match between the NEGF phonon scattering model and Monte Carlo simulations.

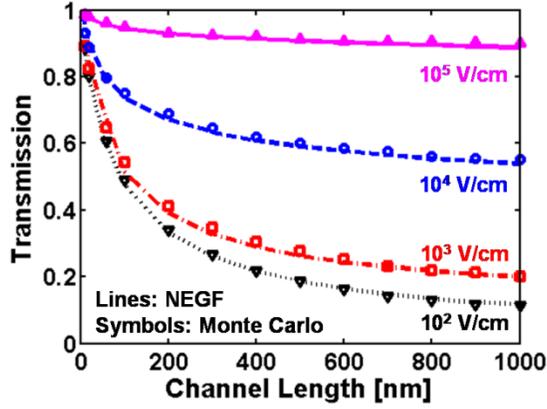

Fig. 4: Plot of transmission vs. channel length with elastic scattering for varying electric fields obtained from Monte Carlo and NEGF simulations. Transmission ~ 1 for fields of $10^6$ V/cm and higher.

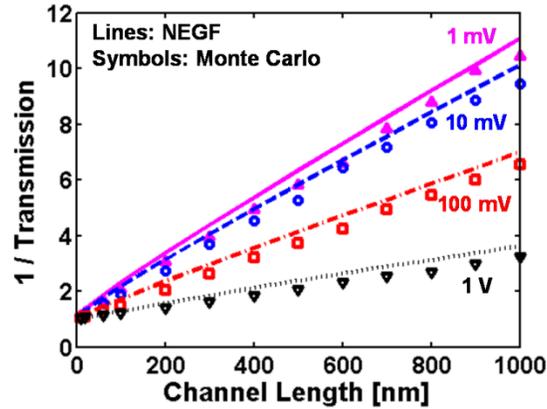

Fig. 5: Plot of transmission inverse vs. channel length for parabolic potential with varying drain bias obtained from Monte Carlo and NEGF simulations with intrasubband acoustic phonons. Mean free path extracted from the slope of low bias (1mV) plot yeilds $\lambda_0$ = 100nm for NEGF and 105nm for Monte Carlo, matching closely the analytical value of $\lambda_0$ = 104nm.

The on-current of a FET is dictated by the charge and velocity at the top of the barrier, hence scattering at the top of the barrier plays a significant role in determining device performance. The NEGF formalism has also been benchmarked using a parabolic potential, which mimics the conduction band profile near the top of the barrier. The conduction band profile is of the form $E_C = -qV_D\, x^2/L^2$, where $V_D$ is the difference between the drain and source conduction band edges and $L$ is the channel length. Figure 5 shows that for small $V_D$, the inverse of transmission varies linearly with channel length as expected $\left(1/T = 1 + L/\lambda_0\right)$. The mean free path extracted via the NEGF ($\lambda_0$ = 100 nm) and Monte Carlo ($\lambda_0$ = 105 nm) are reasonably close to the theoretically expected $\lambda_0$ = 104 nm. It is interesting to note that 1/T varies linearly with L even at large $V_D$. Under high bias condition, the inverse of transmission can be expressed as $1/T = 1 + l/\lambda_0$, where $l$ is the critical length for backscattering. It is believed that for elastic scattering, $l$ is approximately a 'kT length' ($L_{kT}$), which is the distance over which the potential drops by $k_BT/q$ [28]. It has been derived exactly for 1-D transport and is given as $l = L(qV_D/k_BT)\ln(1 + qV_D/k_BT)$ for linear potential and $l = L\sqrt{qV_D/k_BT}\,\tan^{-1}\left(\sqrt{qV_D/k_BT}\right)$ for parabolic potential [24]. $l$ varies linearly with channel length $L$ and monotonically decreases with increase in $V_D$. It is interesting to note that the same linear variation of $l$ with $L$ is observed under high bias in the 2-D transport (fig. 5).

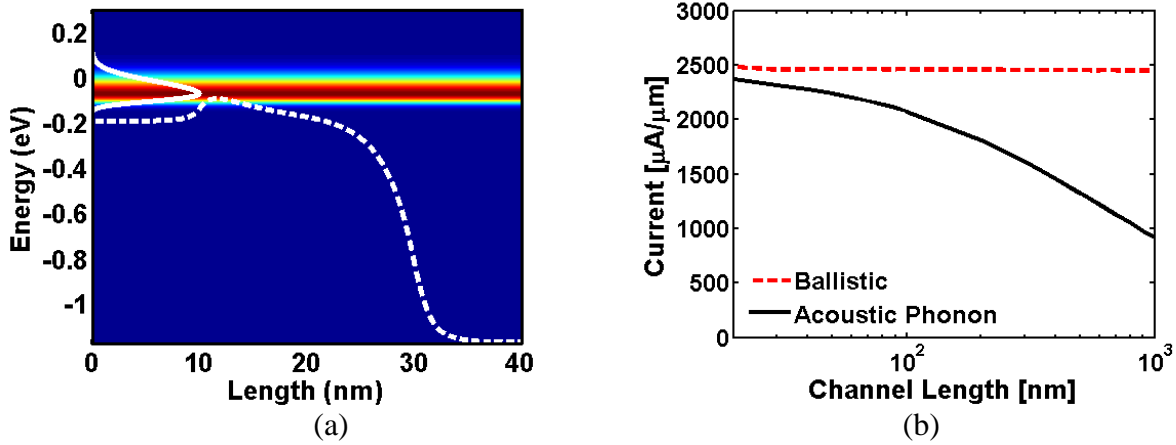

Fig 6: (a) Energy resolved current image with conduction band vs L for a typical MOSFET conduction band profile, and (b) current vs. L from NEGF simulations with ballistic transport and with acoustic phonon scattering.

Finally, acoustic electron-phonon scattering is simulated on a typical MOSFET like conduction band profile. The potential profile (fig. 6a) is self-consistently computed using ballistic NEGF calculations of a double-gate MOSFET structure with 1nm $SiO_2$ insulators and 3nm thick Si channel. The current for 20 nm long channel is close to the ballistic limit, and it steadily degrades for longer channels (fig. 6b), showing that acoustic phonon scattering is not the dominant scattering mechanism for Si MOSFETs with small channel lengths.

*IV.2. Inelastic scattering*

As in the previous case, the inelastic scattering formalism is also benchmarked with Monte Carlo using a single phonon mode. A fixed linear conduction band profile is assumed in the channel region, while the source/drain is assumed to be in equilibrium. Si is used as the channel material, and only the lowest unprimed X valleys are assumed to be occupied. The longitudinal optical (LO) phonon mode with a deformation potential of $11 \times 10^8$ V/cm and phonon energy of 62 meV is used in this study as it is the dominant mode for Si X valley [2].

Figure 7 shows the energy resolved current distribution along the channel and the conduction band profile for an example simulation of inelastic scattering. The deformation potential is artificially increased to highlight the scattering process. Both source and drain injection is considered, with the Fermi levels at 0.2V above the conduction band edge. As there is energy relaxation involved with inelastic scattering, carriers can scatter to higher or lower energies, with higher probability of going to lower energies. At the source end, the injected carriers do not have available states at lower energies, but some scatter to higher energies. Carriers start scattering to lower energies as they progress along the channel. It is interesting to note that due to the high drain bias, drain injected carriers cannot reach the source. These carriers get injected from the drain end at lower energies, get reflected and scattered to higher energies, and go out at the drain contact.

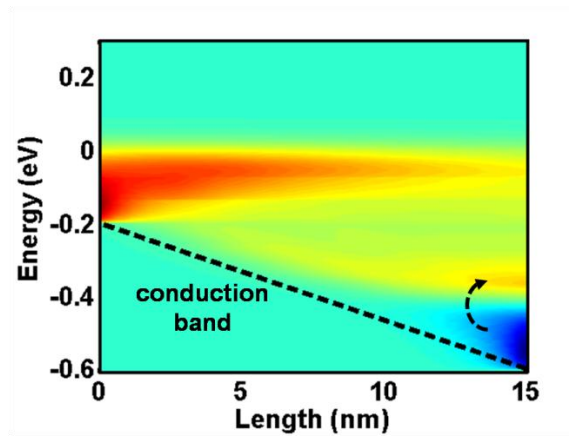

Fig. 7: Energy resolved current density along the channel, along with the conduction band.

Figure 8 plots the transmission as a function of the channel length for high electric fields, obtained from NEGF and Monte Carlo simulations. Very low electric fields are not considered as the carriers don't have empty low-energy states to scatter to. The mean free path is also not extracted as the T in terms of $\lambda_0$ in Sec. IV.1 is valid for elastic process under low bias. The transmission monotonically increases with electric field, and goes to one for fields exceeding $10^6$ V/cm. The transmission is, however, clearly different from the elastic scattering case considered before, where the transmission decreased monotonically with increasing channel length. For the inelastic scattering, transmission decreases with increasing channel length only when the potential drop across the channel is less than the inelastic phonon energy. In this regime, channel carriers can scatter to higher energies only, and they can scatter back and exit at the source end. When the potential drop across the channel is more than the inelastic phonon energy (62 meV in this case), carriers scatter to lower energies, and cannot exit at the source end. In this regime, the current and transmission becomes independent of the channel length. Thus, the transmission becomes constant for $L_{ch} > 620$ nm at $10^3$ V/cm, and $L_{ch} > 62$ nm at $10^4$ V/cm, as confirmed from the transmission plots. This effect is also confirmed from the Monte Carlo simulations, which show a close match to the NEGF results.

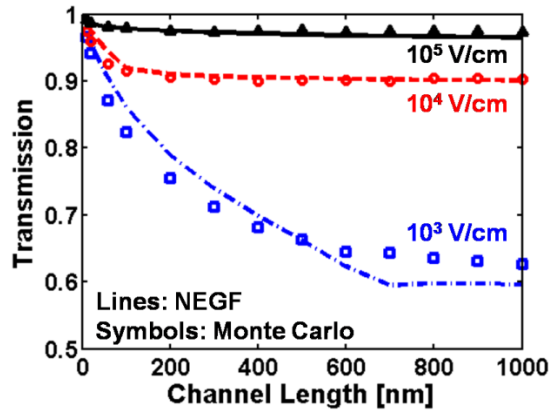

Fig. 8: Plot of transmission vs. channel length with inelastic scattering for linear potential with varying electric fields obtained from Monte Carlo and NEGF simulations. Transmission ~ 1 for fields of $10^6$ V/cm and higher.

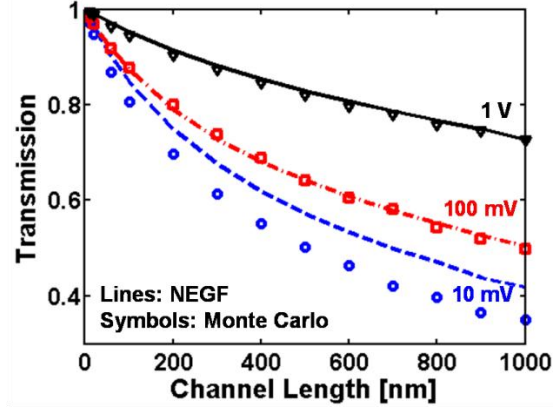

Fig. 9: Plot of transmission vs. channel length with inelastic scattering for parabolic potential with varying drain bias obtained from Monte Carlo and NEGF simulations.

The variation of transmission with $L$ is also benchmarked with Monte Carlo simulations for a parabolic potential, as shown in fig. 9. The characteristics are similar to the linear potential case. When the bias is less than 62 meV, much of optical phonon scattering is suppressed as carriers do not have sufficient states at lower energies to scatter to. The NEGF calculation predicts higher transmissions in such low bias regimes. The discrepancy is larger for low bias. This is caused by the inaccuracy of the approximation we made, that the energy after scattering is independent on the transverse energy. This approximation discounts scattering events with emission of a phonon when the longitudinal energy is less than a phonon energy away from the band edge. In fact such events would occur, if the sum of longitudinal and transverse energy exceeds the phonon energy. That explains why the Monte Carlo method predicts smaller transmission than our version of NEGF. For drain bias greater than 62 meV, the NEGF results agree well with Monte Carlo simulations.

## V. Self-consistent simulations

The transport in the channel region was decoupled from the electrostatics for benchmarking the phonon scattering formalism in the previous section. The scattering formalism is shown to be in close agreement with Monte Carlo simulations for representative potential profiles. However, device simulators must solve the transport self-consistently with the electrostatics. It is imperative to check the consistency of the charge and potential profile with or without scattering. A ultra-thin undoped body silicon-on-insulator (SOI) device with SOI thickness of 8.6nm and channel length of 40 nm is used as an example device structure for the simulations. Similar to the previous benchmarking section, one acoustic and one optical phonon mode is used in these simulations, which correspond to the dominant phonon modes in Si.

The simulation results without and with the effects of scattering are presented in fig. 10. We expect that the presence of scattering should only degrade the current and not the charge, which is governed by electrostatics. This is confirmed by results in fig. 10b and 10c, which show a similar charge at the top-of-the-barrier as a function of gate bias, but lower current due to scattering. The current due to the intrinsic device is shown here, without accounting for the source/drain series resistances. In the ballistic approximation carriers in the channel only occupy forward going k-states (i.e. with momentum towards the drain). Scattering mixes up forward and backward-going states, and the potential energy adjusts to keep the total chargethe same. This explains the slight difference in the potential profiles between ballistic and scattering simulations in fig. 10a.

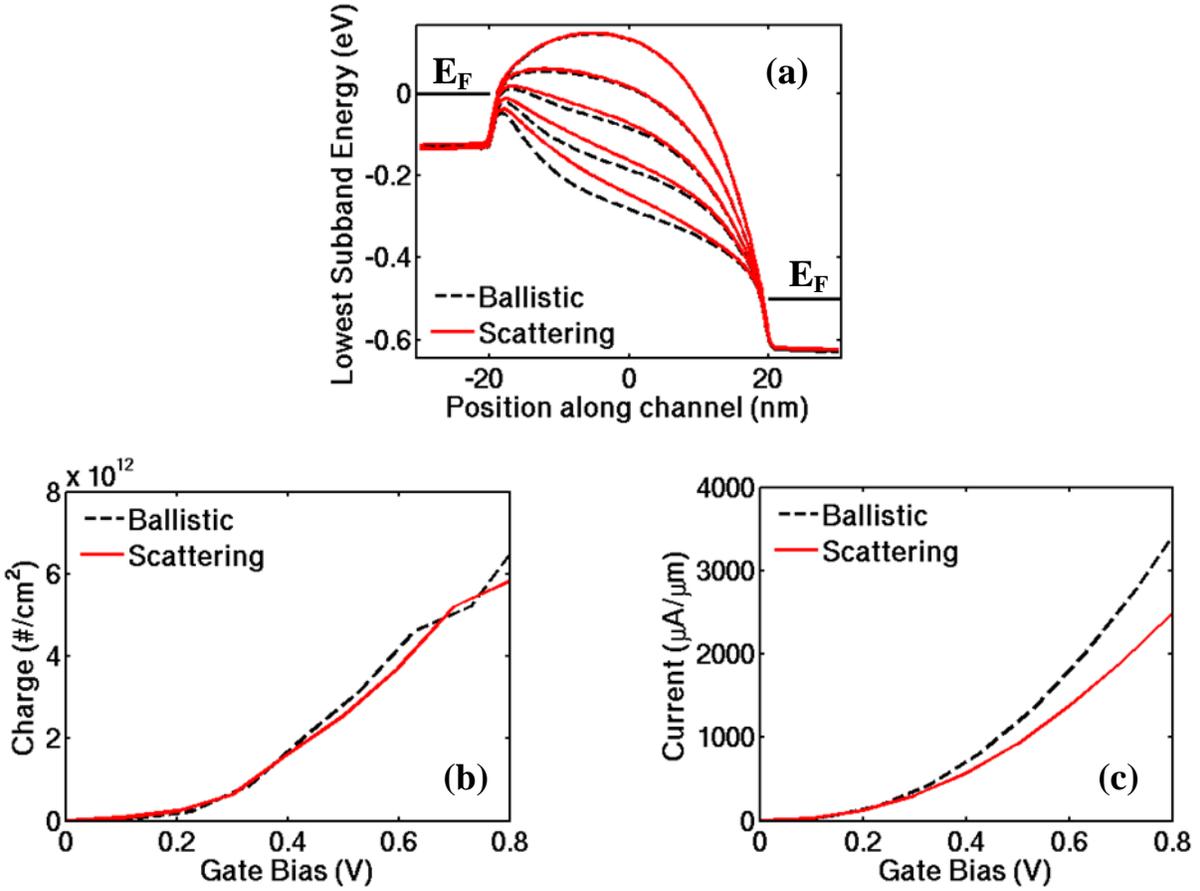

Fig. 10: (a) Plot of the lowest subband energy vs. channel length with gate bias varying from 0 to 0.8V in 0.2V interval, obtained from ballistic NEGF (solid), and NEGF with scattering (dashed). (b) Charge at the top-of-the-barrier and (c) current as a function of gate bias for ballistic and scattering simulations.

**V. Conclusion**

A computationally efficient formalism for incorporating electron-phonon scattering within the NEGF framework is presented. Previously published formalisms have been used for 1-D device simulation of nanowires/nanotubes, but pose significant computational challenge in the case of 2-D transport due to the large number of transverse modes corresponding to the width dimension. The simulation of different transverse modes is avoided in this formalism by implicitly integrating over them in the expressions in/out scattering functions. This novel approach dramatically reduces the computational complexity, making the inclusion of electron-phonon scattering possible in planar MOSFETs. This method is 100~1000 times faster than a full simulation with grid in the large transverse width dimension. This formalism is incorporated in the widely used NEGF simulator nanoMOS, and tested for optical and acoustic phonons in non-polar materials, which are coupled to electrons via deformation potential interactions. The in/out scattering functions and the self energy involve only local interaction in these cases, resulting in diagonal matrices and allowing an exact solution without full matrix inversion. The NEGF-scattering framework is rigorously benchmarked against classical Monte Carlo simulations for model channel structures where quantum effects will be minimal, with a broad range of channel lengths and biasing conditions. The results show an extremely good match between the two methods except for small channel lengths, where contacts dominate the device behavior, which is not captured by the classical Monte Carlo technique. The scattering formalism is tested within a self-consistent framework for a realistic device structure, and is shown to degrade the current when compared to ballistic simulations.

## Acknowledgements

This work is supported by SRC grant 1871.001, with computational support from the Network for Computational Nanotechnology. The authors wish to thank Supriyo Datta for stimulating discussions.